\documentclass[twocolumn,showpacs,amssymb,aps]{revtex4}
\usepackage{amsmath}
\usepackage{amsfonts}
\usepackage{amssymb}
\usepackage{graphicx}
\usepackage{epsfig}
\usepackage{dcolumn}
\usepackage{bm}

\newcommand{\req}[1]{Eq.\,(\ref{#1})}

\def\lessim{\lower.5ex\hbox{$\; \buildrel < \over \sim \;$}}

\begin{document} \hbadness=10000
\topmargin -1.cm
\preprint{}

\title{Unstable Hadrons  in Hot Hadron Gas: in Laboratory, and in the Early Universe}
\author{Inga Kuznetsova and Johann Rafelski}
\affiliation{Department of Physics, University of Arizona, Tucson, Arizona, 85721, USA}

\begin{abstract}
We study kinetic master equations for chemical reactions 
involving the formation and the natural decay of unstable
particles in a thermal bath. We consider the decay channel of one 
into two particles, and the inverse process, fusion of two 
thermal particles into one. We present the master 
equations for the evolution of the density of the unstable 
particles  in the early Universe. 
We obtain the thermal invariant reaction rate
using as an input the free space (vacuum) decay time
and show the medium quantum 
effects on $\pi+\pi \leftrightarrow \rho$ reaction relaxation time.
As another laboratory example we describe the
$K+K \leftrightarrow \phi$ process in thermal hadronic 
gas in heavy ions collisions. A particularly interesting application of our
formalism   is the  $\pi^{0}\leftrightarrow \gamma +\gamma$ process
in the early Universe. We also explore the physics of 
$\pi^{\pm}$ and $\mu^{\pm}$ freeze-out in the Universe. 
\end{abstract}

\date{January, 28, 2009}

\pacs{95.30.Cq, 52.27.Ny, 24.10.Pa}

\maketitle

\section{Overview}

\subsection{Particles in the Universe for $T>5$~MeV}
This study began with the question: At what temperature
 in the expanding early Universe does the reaction
\begin{equation}
\pi ^{0}\leftrightarrow \gamma +\gamma  \label{pigg}
\end{equation}%
`freeze' out, that is the $\pi^0$ decay overwhelms the production rate and the yield
falls away from chemical equilibrium yield. Because the $\pi^0$ life span (8.4$\,10^{-17}$ s) is
rather short, one is tempted to presume that the decay process (arrow to the right) dominates. However, 
there must be a detailed balance in the thermal bath: the production process (arrow to the left) 
in a suitable environment must be 
able to form $\pi^0$ with strength corresponding to the decay process lifespan. 

We demonstrate here that the $\pi^0$ production and equilibration relaxation time 
is of the same order of magnitude as the lifespan of $\pi^0$ in the 
post-quark-gluon-plasma hadronization Universe, $T<200$ MeV.
The point is that the $\pi^0$ life span is much shorter than the Universe expansion time 
(inverse expansion rate) $1/H$~\cite{Kolb:1988aj}:
\begin{equation}
H=\frac{\dot R}{R}=1.66 \sqrt{g^*}\frac{T^2}{m_{pl}},
\end{equation}
where $g^*$ is the number of degrees of freedom. $m_{pl}=1.2211\,10^{19}$ GeV is the Plank mass.
Figure \ref{taupi01} 
compares the $\pi^0$ production-equilibration time (blue solid line) with  the Universe
expansion time  $1/H$ {dashed (green) line]. We see that $\pi^0$ equilibration time is much 
shorter, by 14 orders of magnitude at $T=10$ MeV, compared to the Universe expansion time constant.

\begin{figure}[tbp]
\centering \includegraphics[width=8.6cm,height = 8.5cm]{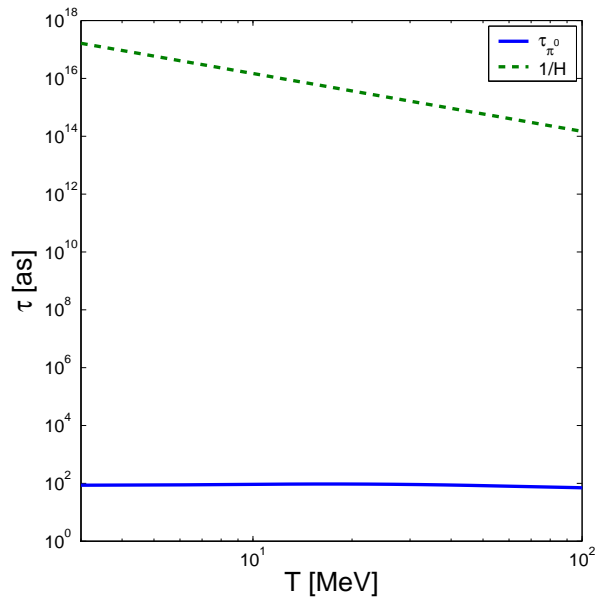}
\caption{{\protect\small {(Color online) $\pi^0$ equilibration time [solid (blue) line] 
and Universe expansion time $1/H$ as  functions of temperature [dashed (green) line].}}} \label{taupi01}
\end{figure}

The  reason for this is that in thermal equilibrium  the photon  density remains high 
[dash-dotted  (red) line in figure \ref{npi2}] also for relatively small $T$. 
Thus there is a small  non-negligible
probability of finding high energy photons capable
to produce   $\pi^0$, whose density  at low $T$ is very
small (solid blue line in figure \ref{npi2}).  The $\pi^0$ 
production has   enough time to equilibrate with the decay process. 
Therefore the $\pi^0$ density does not freeze out, but
decreases with decreasing ambient temperature of the expanding Universe,
all the time remaining in chemical equilibrium with the photon abundance.

\begin{figure}[tbp]
\centering \includegraphics[width=8.6cm,height = 8.5cm]{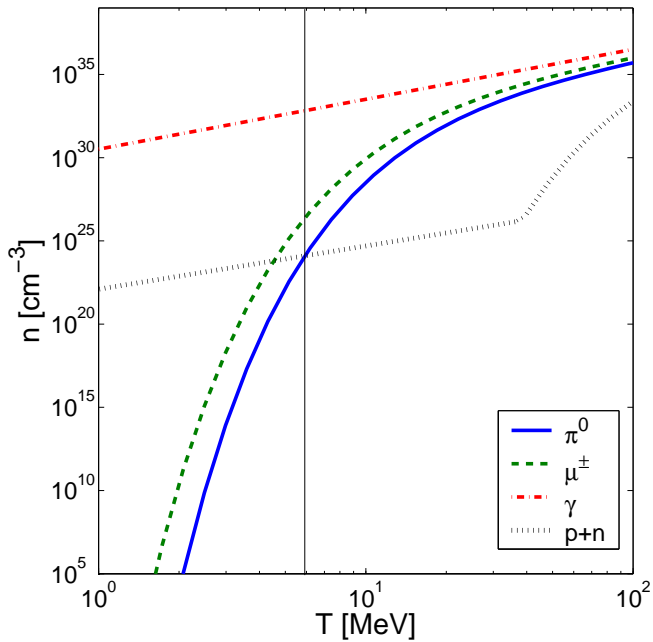}
\caption{{\protect\small {(Color online) Thermal equilibrium density as a functions of temperature for:
$\gamma$ [dash-dot (red) line], 
$\pi^0$ [solid (blue) line], 
$\mu^{\pm}$ pair [dashed (green) line], 
and nucleons $p+n$ [dotted (black) line~\cite{FromerthRafelski}].}}} \label{npi2}
\end{figure}

Let us recall how the Bose distribution describes $\pi^0$-density:
\begin{equation}
n^{eq}_{\pi^0} =\int \frac{d^{3}p}{(2\pi)^{3}} \frac{1}{e^{u\cdot p/kT}-1},  \label{npi} 
\end{equation}
Here $u^\mu$ is the four-velocity of the observer with 
reference to the heat bath rest frame, $p^\mu$ is momentum four-vector
\begin{equation}
\label{4p}
p^\mu=\left(\frac{E}{c}, \vec p\right),\qquad  E=\sqrt{\vec p^{\,2}+m^2}
\end{equation}
of the particle considered: a similar expression applies for photons, which have 
two fold spin degeneracy, and $m\to 0$.  The resulting  $\pi^0$ density
falls exponentially when $kT<m_{\pi^0}c^2$ (henceforth units are chosen 
such that $\hbar=c=k=1$).  However this density remains high 
compared to the nucleon's density in the Universe 
(dotted (black) line in Fig. \ref{npi2}, taken from \cite{FromerthRafelski}), 
down to a temperature of about 6 MeV. This is the lower $T$-limit of validity in our
present study, as we consider particle production reactions in a 
particle-antiparticle symmetric Universe. 

Some of the results we derive here were presented in \cite{Kuznetsova:2008jt} 
without a derivation: there we considered a laboratory $e^+e^-\gamma$ plasma and 
postponed the theoretical and analytical details. Here we  evaluate 
the reaction relaxation time  for reactions involving two particles fusing into one particle,
and/or particle decay into two,  \req{pigg}, and relate this to the lifespan of decaying 
particle in vacuum.  To complement this in chapter \ref{4} subsection (c) we also consider
$\pi^{\pm}$, which can be  equilibrated by the reaction:
\begin{equation}
\pi^0+\pi^0 \leftrightarrow \pi^+ + \pi^- \label{pipm}.
\end{equation}
and also by reactions involving muons
\begin{equation}
\pi^\pm \leftrightarrow \mu^\pm +\nu_{\mu}(\bar{\nu}_{\mu}), \label{pimunu}
\end{equation}

We also consider how fast muons are produced in the reactions:
\begin{equation}
\gamma+\gamma \leftrightarrow \mu^+ + \mu^-,\,\,\,\,\,\,\,e^+ + e^- \leftrightarrow \mu^+ + \mu^-; \label{muprod}
\end{equation}
and show that these particles also do not freeze out down to 
a $T$ value of a few MeV~\cite{Kuznetsova:2008jt}. The muon density is slightly higher
than that pions because of their smaller mass; see the dashed (green) line in Fig. \ref{npi2}.
Another reaction that may influence muon chemical equilibration 
is the decay of one particle to three particles, 
and the reverse  reaction, including neutrinos:
\begin{equation}
\mu^{\pm} \leftrightarrow e^{\pm} + \nu_{e}(\bar\nu_e) + \bar\nu_{\mu}(\nu_{\mu}).\label{munu}
\end{equation}
However in this case the exact influence  of medium effects on the reaction rate is more complicated and 
we do not consider this reaction  in complete detail here. We do, however, compare 
the relaxation times  of particle production in all mentioned reactions with the 
Universe expansion rate  to see if the particle densities stay in chemical equilibrium.  

\subsection{Degrees of Freedom in the Universe}

The lifespans of all unstable hadrons and leptons, except for neutrons $n$, 
are much shorter than the Universe expansion rate for $5<T<200$ MeV.  
Here we show that, as a result, all unstable particles stay in chemical equilibrium,
including neutrons which are effectively stable on the time scale of expansion.
The importance of this remark is that we can evaluate the active effective 
degeneracy (degrees of freedom) in 
the Universe in the entire  temperature domain including 
all unstable hadron states. In the hadron phase we define 
the effective degeneracy using as reference the Stephan-Boltzmann law,
\begin{equation}
g_E(T)=\frac{\epsilon}{\sigma T^4},\qquad\sigma=\frac{\pi^2}{30},\label{gEdef}
\end{equation}
where $\epsilon$ is the energy density:
\begin{equation}
{\epsilon} = \int \sum_i g_iE_if_i(p)d^3p, \quad E_i=\sqrt{m_i^2+\vec p^{\,2}}, 
\end{equation}
whith the sum over all particles present. 

In Fig.~\ref{degen}
we show  the degeneracy \req{gEdef} as a function of $T$. 
The dashed (red) line  accounts for the photon, 
three families of neutrino and antineutrino, 
electron, positron,  muon, and antimuon contributions. 
The dot-dashed (green) line adds pions; the solid (blue) line,   
all hadrons.  Pions begin to contribute 
noticeably  to degeneracy at $T> 30$~MeV. Among hadrons we included all light 
and strange mesons and baryons, up to a mass of about 1700 MeV. The finite 
density of $p$ and $n$ is also included. As noted earlier, this is 
a more complicated case; fortunately the  finite baryon density 
contributes  at just at a few  percent   to  $g_{E}$ near the hadronization 
temperature, where the particle-antiparticle symmetry is good to 10 orders of magnitude. 
\begin{figure}[tbp]
\centering \includegraphics[width=8.6cm,height = 8.5cm]{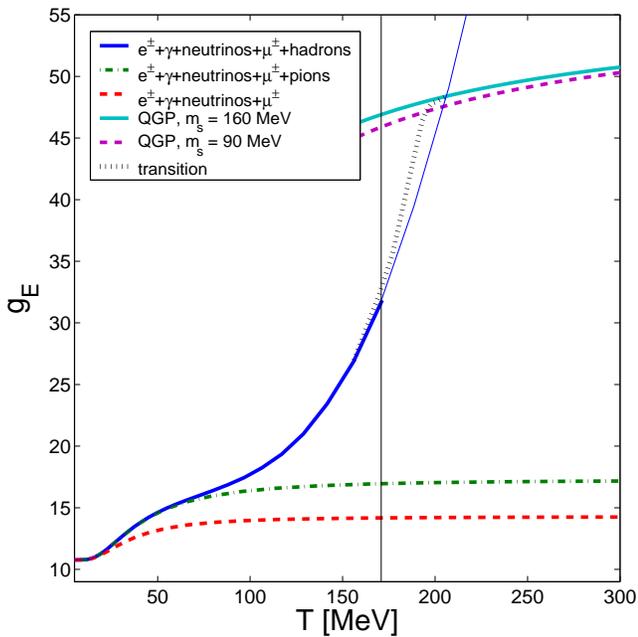}
\caption{{\protect\small {(Color online) Effective degeneracy $g_{E}$ in the Universe based
on the energy density of hadrons, and for QGP, as a function of $T$. 
See text for more details.}}} \label{degen}
\end{figure}

The boundary between quark-gluon and  hadron phase (vertical line) is near
$T_h=170$ MeV. In figure~\ref{degen} we also show degeneracy in 
QGP for $T>160$~MeV (upper lines). The results
shown are based on our earlier detailed study of QGP 
properties~\cite{Kuznetsova:2006bh}. Here,  the QGP degeneracy is shown for the
extreme cases of either  strange quark $m_s= 90$~MeV [dashed (purple) line] or  160~MeV 
[solid (turquoise) line]. Since the expansion of the Universe is relatively slow compared to 
expansion of QGP in laboratory, heavy and strange quarks also have enough time  
to reach chemical  equilibrium density in the QGP temperature range presented in graph. 

We see in figure~\ref{degen} that the effective degeneracy of hadrons, 
while rising fast, is still  smaller than the degeneracy in QGP in the domain 
of phase transformation temperature near 160-170 MeV. Many heavy hadron states 
may be missing from the experimental tables. Even though their individual 
contribution to the degeneracy is decreasing, their number is expected 
to grow rapidly, in accordance with the Hagedorn hypothesis, in which hadron  mass
spectrum diverges exponentially near the hadronization temperature.  
This theoretical exponentially growing component leads to a smoother 
transition between hadronic gas and QGP, as is qualitatively indicated in Fig.\ref{degen} [dotted (black) line].  

\subsection{Production and decay of unstable particles}

We show here for the first time the 
detailed derivation of relaxation time for reactions involving 
one-to-two particles  in the thermal medium, which we considered 
in ~\cite{Kuznetsova:2008zr} - \cite{Kuznetsova:2006bh}.
In the rest frame of the decaying particle $m_{3}$, the reaction
\begin{equation}
A_1+B_2 \leftrightarrow C_3  \label{123} 
\end{equation}%
requires that  $m_1 + m_2 \le m_{3}$, which allows the spontaneous decay process. This
is easily seen considering
\begin{eqnarray}
m_3^2&=&(p_1+p_2)^2\nonumber\\
&=&(m_1 +m_2)^2 +2(E_1E_2-m_1m_2-\vec p_1\cdot \vec p_2)\nonumber\\
&\ge& (m_1 +m_2)^2. \label{123c}
\end{eqnarray}
In the last inequality we used $E_1^2E_2^2\ge (m_1m_2+\vec p_1\cdot \vec p_2)^2 $ 
which can be reorganized to read  
$(m_1 \vec p_2 -m_2 \vec p_1)^2\ge  \vec p_1\cdot \vec p_2- \vec p_1^{\,2} \vec p_2^{\,2}$.
This is always true since the right hand side is always negative, or zero if both 
vectors are parallel. The equality  sign corresponds to the case $m_1+m_2=m_3$, 
where the reaction rate vanishes by virtue of vanishing phase space.
This text-book exercise shows that the reaction Eq.\,(\ref {pigg}) is 
possible when condition \req{123c} is satisfied.  

The constraint  Eq. (\ref{123c}) forbids many reactions. For example, the hydrogen formation
 $p+e\to $H is forbidden, as for a bound state, $m_H<m_p+m_e$. Thus
there must be  a  second particle in the final state.
The  electron capture involves  either a radiative emission,
$p+e\to $H$+\gamma$, or a surface/third atom, which picks  the recoil momentum.
The situation would be different if there were "resonant"' intermediate states of relatively long life span
with energies above the ionization threshold.  Such "doorway" resonances are available in many important
physical processes.

It is natural to  evaluate the rates of the processes of interest, Eq (\req{123})
in the rest frame of particle `3',  boosting, as appropriate,
from or to laboratory frame. To do this effectively  we  need the master population
equations in an explicitly covariant fashion, which is discussed in Sec: \ref{2},
see Ref.~\cite{Kuznetsova:2008zr}. 
The kinetic equation for time evolution of number
$N$ of decaying particles $3$ can be written as
\begin{equation}
\frac{1}{V}\frac{dN_{3}}{dt}=\left( \frac{\Upsilon _{1}\Upsilon _{2}}{\Upsilon_{3}} -1\right)
               \frac{dW_{3\rightarrow 12}}{dVdt},  \label{fe}
\end{equation}%
where ${dW_{3\rightarrow 12}}/{dVdt}$ is the decay rate of particle $3$ and
$\Upsilon _{i}$ is the fugacity of particle $i$. Here the number density
$n_{i}$ of particle $i$ in thermal (kinetic), but not necessarily in
chemical, equilibrium is given by
\begin{eqnarray}
\frac{N_i}{V} \equiv n_{i} &=&\frac{1}{(2\pi )^{3}}\int d^{3}p_{i}f_{b/f}(p_{i}), \label{nf} \\
f_{b/f}(\Upsilon _{i},p_{i}) &=&
    \frac{1}{\Upsilon _{i}^{-1}e^{(u\cdot p_{i}-\mu_i)/T}\mp 1}.  \label{f}
\end{eqnarray}%
$f$ is the covariant form of the usual Bose or Fermi distribution
function defined in the rest frame of the thermal bath, and describes the
corresponding quantity in a general reference frame where the thermal bath
has the relative velocity defined by $u^{\mu }$. In the rest frame of the 
thermal bath frame we have:
\begin{equation}
u^{\mu }\rightarrow \left( 1,\vec{0}\right) .  \label{4v-rest}
\end{equation}
$p_{i}$ is the four-momentum-vector of particle $i$: 
\begin{equation}
p_{i}^{\mu }=\left( E_{i},\vec{p_{i}}\right) .  \label{4v}
\end{equation}%
$\mu_i$ is the chemical potential, which shows the asymmetry in particle and 
antiparticle densities $\mu_{i}=-\bar \mu_{i}$. For reactions considered here 
we have $\mu_i \simeq  0$. This was assumed in Eq.\ref{fe}. Note that the
distribution function $f$ is a Lorentz scalar but the spatial density $n_{i}$ is
not. 

Particle $C_3$ attains the chemical equilibrium when the following
condition among fugacities is satisfied:
\begin{equation}
\Upsilon_{1}\Upsilon_{2}=\Upsilon_{3}.  \label{equilcon}
\end{equation}%
This, as expected, is equivalent to the Gibbs condition for the chemical
equilibrium. In Sec. \ref{3} we evaluate the invariant rate using the vacuum decay time  
established  in  the rest frame of the decaying particle, and we 
discuss the behavior of the average
decay rate of an unstable particle in the presence of the thermal bath. In
Sec. \ref{4}, we apply our formalism to two examples: \\
\indent {\bf a)} We study the formation
and decay rate of the $\rho $ meson through $\pi +\pi \leftrightarrow \rho $ in
a baryon-free hot hadronic gas, where mesons are considered in thermal and
chemical equilibrium. \\
\indent {\bf b)} We  consider the  reaction $\gamma +\gamma \leftrightarrow \pi^{0}$ 
in the early Universe and find that the expansion of the
Universe is slow compared to pion equilibration, which somewhat
surprisingly (for us) implies that  $\pi ^{0}$ is at all times in
chemical equilibrium (but at sufficiently low temperatures e.g.  3-4 MeV, the
local density of  $\pi^{0}$ is too low to apply the methods of
statistical physics).\\
\indent {\bf c)} We consider the reaction~Eq.(\ref{pimunu}) as an example of the decay 
of  $\pi ^{\pm}$ to fermions, and the reverse reaction, and show that $\pi^\pm$ and $\mu^\pm$ 
are also in chemical equilibrium until their equilibrium density vanishes 
at low temperatures (about 3-4 MeV) because of the large mass. Also, we discuss
neutrinos equilibration by way of this reaction.\\
\indent {\bf d)} We study $\phi$ mesons evolution considering the reaction $K+K \leftrightarrow \phi$ 
in thermal hadronic gas in heavy-ion collisions. 

To conclude this overview we draw attention to the fact 
that unlike the one-to-two reaction the two-to-two
reactions 
\begin{equation}
A_1+B_2 \leftrightarrow C_3 +D_4 \label{1234}
\end{equation}%
have been extensively studied in the past,  in the context
of astrophysics and cosmology~ \cite{xx,Kolb:1988aj}
and heavy ion reactions~\cite{hadBook}.  However, the simpler one-to-two situation
has escaped attention so far, and the adaptation of kinetic methods 
is in detail not trivial given the
 novel quantum and relativistic effects involving  particle decay.
Aside from cosmology implications, the
insights gained in this study are clearly of
relevance to the general understanding of QGP and
hadron gas evolution in relativistic heavy ion collisions. For
example our present work allows to consider the chemical yields arising in
reactions such as $\rho \leftrightarrow \pi \pi $, $\pi^{0} \leftrightarrow \gamma \gamma $, 
$\Delta \leftrightarrow N\pi $ and $K+K \leftrightarrow \phi$~\cite{Kuznetsova:2008zr,Kuznetsova:2008hb}.


\section{Kinetic equations for decaying particles}\label{2}

\subsection{Decaying particle density evolution equation}
Consider an unstable particle, say $C_3$, which decays   into other two
particles,
\begin{equation}
C_3 \rightarrow A_1+B_2  \label{2eq}
\end{equation}
in the vacuum. In a dense and high-temperature thermal ambient  phase,   particles
$A_1 $ and $B_2$ are present, and the inverse reaction,
\begin{equation}
A_1+B_2\rightarrow C_3  \label{1eq}
\end{equation}
can occur, producing the particle we called $C_3$. For now we assume that the abundance of
particle $C_3$ changes solely by decay \req{2eq} and (thermal) 
production \req{1eq} reactions. The
time variation of the number of particles $N_3$ than is controlled by the master equation
\begin{equation}
\frac{1}{V}\frac{dN_{3}}{dt}=
    \frac{dW_{12\rightarrow 3}}{dVdt}
    -\frac{dW_{3\rightarrow 12}}{dVdt},  \label{popeq}
\end{equation}
where $dW_{12\rightarrow 3}/dVdt$ is the production rate per unit volume of
particle type $C_3$ via Eq.(\ref{1eq}) and $dW_{3\rightarrow 12}/dVdt$ is the decay rate of
particle type $C_3$ per unit volume.

A very similar master equation controls the abundance of particle $A_1$ and $B_2$ 
\begin{equation}
\frac{1}{V}\frac{dN_{1,2}}{dt}=\frac{dW_{3\rightarrow 12}}{dVdt}
                    -\frac{dW_{12\rightarrow 3}}{dVdt} +R_{\rm other},  \label{popeq23}
\end{equation}
where the rate $R_{\rm other}$ is caused by other reactions influencing the 
abundance of particles of type $A_1$ and $B_2$. 

As an example, consider
the reaction $\rho \leftrightarrow \pi \pi $ in dense hot matter formed 
in heavy-ion collisions. Pions can be easily created by inelastic 
collisions of other hadrons and thus  we have to deal with a multicomponent 
system involving  $R_{\rm other}$ when looking at $\pi$ abundance, 
but to evaluate $\rho$ abundance the dominant terms are as in Eq.(\req{popeq}).
We often can assume that $R_{\rm other}$ dominates the yield gains and losses 
and thus we can use the thermal distribution for the 
particles  $A_1$ and $B_2$, which in the example above are pions.

In the following, we thus assume that particles  $A_1$ and $B_2$ are in thermal 
equilibrium and further, we assume that the system is spatially homogeneous.
In a thermal equilibrium, the
dynamical information can be obtained from the single particle distribution
function $f\left( p\right) $ for each particle, see \req{f}. $f$ is controlled 
by two parameters, the temperature $T$ 
and the fugacity $\Upsilon $ . In this paper, we assume that  the fugacity
$\Upsilon $ changes over time by way of chemical reactions much more rapidly than
does the temperature $T$ of the ambient thermal bath, and thus we can consider 
reactions at a given constant $T$. This assumption is certainly valid in the
domain of temperatures we consider, and may fail only at the very highest
primordial $T$ in the early Universe. 

\subsection{Decay and production rates}
The thermal production rate ${dW_{12\rightarrow 3}}/{dVdt}$ and the decay
rate of particle $3$ under the thermal background
${dW_{3\rightarrow 12}}/{dVdt}$ can then be expressed using these distribution 
functions for each of the particles involved in the reaction. 
According to the boson or fermion nature of the particle $A_1$, we have to
consider different cases. If particle $A_1$ is a boson, then there are two
cases of the decay and production mode, and if the particle $C_3$ 
is a fermion, it can only  decay into a boson and a fermion:
\begin{eqnarray}
&&\mathrm{boson_{3}\longleftrightarrow boson_{1}+boson_{2},} \\
&&\mathrm{boson_{3}\longleftrightarrow fermion_{1}+\overline{fermion_{2}}.}\\
&&\mathrm{fermion_{3}\longleftrightarrow boson_{1}+fermion_{2}.}
\end{eqnarray}%
 
Accordingly, the Lorentz invariant transition 
probability per unit time and unit volume
corresponding to the process Eq.(\ref{123}) is
\begin{widetext}
\begin{eqnarray}
\frac{dW_{12\rightarrow 3}}{dVdt} &=&\frac{1}{1+I}
  \frac{g_{1}}{(2\pi )^{3}}\int \frac{d^{3}p_{1}}{2E_{1}}f_{b,f}(\Upsilon _{1},p_{1})
 \frac{g_{2}}{(2\pi )^{3}}\int \frac{d^{3}p_{2}}{2E_{2}}f_{b,f}(\Upsilon _{2},p_{2})
          \int\frac{d^{3}p_{3}}{2E_{3}\left( 2\pi \right) ^{3}} \notag \\[0.2cm]
&&\times  
\left( 2\pi \right) ^{4}\delta ^{4}\left( p_{1}+p_{2}-p_{3}\right)
\frac{1}{g_{1}g_{2}}\sum_{\mathrm{spin}}\left\vert \langle p_{1}p_{2}
 \left\vert M\right\vert p_{3}\rangle \right\vert ^{2}
   \left( 1\pm f_{b,f}(\Upsilon_{3},p_{3})\right) ,  \label{pp}
\end{eqnarray}%
\end{widetext}
where $I=1$ for the case of a reaction between two indistinguishable 
particles $A_1$ and $A_2$, and $I=0$ if
 $A_1$ and $B_2$ are distinguishable. 
The factor $1/(g_{1}g_{2})$ and the summation are
caused by averaging over all initial (iso)spin states. The last factor in \req{pp} 
accounts for the enhancement or hindrance of the final-state phase owing to the quantum
statistical effect, as  introduced first by Uehling and Uhlenbeck~\cite{BUU}. The
upper sign $^{\prime }+^{\prime }$ is for the case when particle $C_3$ is a boson;
the lower sign $^{\prime }-^{\prime }$ when it is a fermion.  Eq. (\ref{pp})
is manifestly Lorentz invariant and  therefore it can be used in any frame of reference.

This rate is related  by a detailed balance relation~\cite{Kuznetsova:2008zr} 
to the particle $C_3$ decay rate
\begin{equation}
\frac{dW_{12\rightarrow 3}}{dVdt}\Upsilon_{3}=
\frac{dW_{3\rightarrow 12}}{dVdt} \Upsilon_{1}\Upsilon_{2}.\label{pdr}
\end{equation} 
Therefore chemical equilibrium
${\Upsilon_{1}\Upsilon _{2}}={\Upsilon _{3}}$ corresponds to the condition of 
equal decay and production rates as we expected. Using  \req{pdr}, 
\req{popeq}  can be written in the form of Eq.(\req{fe}).

Equation (\ref{fe}) can be further simplified  by defining 
the decay time in matter of particle $i$:
\begin{equation}
\tau _{i}=\frac{dn_{i}/d\Upsilon _{i}}{R}, \label{taui}
\end{equation}
where rate:
\begin{equation}
R=\frac{1}{\Upsilon _{3}}\frac{dW_{3\rightarrow 12}}{dVdt}
  =\frac{1}{\Upsilon _{1}\Upsilon _{2}}\frac{dW_{12\rightarrow 3}}{dVdt}. \label{A}
\end{equation}
We show in the next section that this definition has the right vacuum limit and that the dynamical 
equations assume a particularly simple form. However, the reader should observe that
other definitions could be considered. 

It is convenient to 
introduce kinematic reaction times in analogy to the dynamic  expression \req{taui}.  Doing this 
 we cast \req{fe} into the form of an equation for $\Upsilon_3$
\begin{equation}
\dot \Upsilon_{3}=\frac{1}{\tau_T}\Upsilon_3+\frac{1}{\tau_S}\Upsilon_3
  + \frac{1}{\tau_3}(\Upsilon_1\Upsilon_2 - \Upsilon_{3}), \label{uppiu}
\end{equation}
where we defined the kinematic relaxation times related to the evolution of temperature and entropy
\begin{eqnarray}
&&\frac{1}{\tau_T}\equiv -T^3g^*\frac{d (n_{\pi}/(\Upsilon_3
g^*T^3))/dT}{dn_{\pi}/d{\Upsilon_3}}{\dot T},\label{tauT} \\
&&\frac{1}{\tau_{S}}\equiv
-\frac{n_{\pi}/\Upsilon_3}{dn_{\pi}/d{\Upsilon_3}}\frac{d\ln (g^*VT^3)}{dT}
\dot{T}. \label{Seq}
\end{eqnarray}
We introduced the minus sign above in order to have $\tau_T$, $\tau_S>0$. Compared to our 
earlier definition~\cite{Kuznetsova:2008zr}, we now included $g^*$ in  $\tau_T$ and $\tau_S$. 

While in principle the values of $\tau_T$ and $\tau_S$ are unrelated, for a given kinematic
stage of system evolution   the temperature change can be related to  entropy change. For example, for the 
radiation-dominated epoch of the Universe we have
\begin{equation}
\frac{\dot T}{T}=-\frac{ \dot R}{R}. \label{Tch}
\end{equation}
In the radiation-dominated Universe the entropy conservation further implies that
\begin{equation}
\frac{1}{\tau_S} \rightarrow 0.
\end{equation}
Freeze-out from chemical equilibrium arises for 
\begin{equation} 
{\tau_{T}}\approx \tau_3 \label{frcon}
\end{equation}
When $\tau_{T}$ is smaller than $\tau_3$, that is the kinematic term in \req{uppiu} 
is important,   $\Upsilon_3$ begins 
to increase often rapidly, as the number of particles `3' is preserved  but their 
density decreases owing to dilution in expansion. Because $\Upsilon_3 >1 $  the multiplicity 
of particle `3' in  the slow decay  is dominant, especially so when the
particles 1 and 2 yields remain in chemical  equilibrium by the action of 
other processes. These considerations can be of great importance  in the 
study of  dark-matter particle abundance where
the lifespan against decay and/or annihilation is comparable to the life span of the 
Universe. We postpone further discussion to a more appropriate opportunity.

\section{Evaluation of the invariant decay (production) rate}\label{3}

\subsection{General case}

The vacuum decay width of particle  $C_3$ in its own rest frame can be found in   
textbooks. In our notation,
\begin{widetext}
\begin{align}
\frac{1}{\tau _{0}}& =\frac{1}{2m_{3}}\frac{1}{1+I}
\int \frac{d^{3}p_{1}}{2E_{1}\left( 2\pi \right) ^{3}}
\int \frac{d^{3}p_{2}}{2E_{2}\left( 2\pi \right)^{3}}
\left( 2\pi \right) ^{4}\delta ^{4}\left( p_{1}+p_{2}-p_{3}\right)\frac{1}{g_3}
\sum_{spin}\left\vert \langle p_{1}p_{2}\left\vert M\right\vert p_{3}\rangle \right\vert ^{2}  \notag \\
& =\frac{1}{2m_{3}g_3}\frac{1}{4\left( I+1\right) \left( 2\pi \right) ^{2}}
\int\frac{d^{3}p}{E_{1}E_{2}}\delta (E_{1}+E_{2}-m_{3})\sum_{spin}\left\vert
\langle \vec{p},-\vec{p}\left\vert M\right\vert m_{3}\rangle \right\vert ^{2}
=\frac{1}{8m_{3}^2g_3}\frac{p}{(I+1)\pi}\sum_{spin}\left\vert
\langle \vec{p},-\vec{p}\left\vert M\right\vert m_{3}\rangle \right\vert ^{2} \label{VacTau}
\end{align}
\end{widetext}
Here $p=p_1=p_2$ and $E_{1,2} =\sqrt{p^2+m^2_{1,2}}$
are, respectively, the magnitudes of the momentum and the energies, 
of the two particles $A_1$ and $B_2$ in the
rest frame of particle $C_3$
\begin{eqnarray}
E_{1,2}&=&\frac{m_{3}^{2}\pm (m_{1}^{2}-m_{2}^{2})}{2m_{3}}, \nonumber \\
\vec p^{\,2}&=& 
\frac{m_3^2}{  4} -\frac{m_1^2+m_2^2}{2}+\frac{(m_1^2-m_2^2)^2}{4m_3^2}. \label{encon}
\end{eqnarray}
The magnitude of three-momentum $|\vec p|$ 
is of course the same for particle $A_1$ and $B_2$ in the
rest frame of decaying particle $C_3$.

We denote  by $\tau_3^{\prime }$ the decay rate of particle $C_3$ 
in the  rest frame of   the
thermal bath in which it is emerged,  $E_{3}$ and $p_{3}$ are  
the corresponding energy and the momentum of particle $C_3$ which changes
with thermal velocity distribution.  The thermal decay reaction rate per unit 
volume $dW_{3\rightarrow 1+2}/dVdt$ is then obtained by  weighting   $1/\tau_3^{\prime }$
with the probability to find the particle at a given momentum and introducing the 
Lorentz factor $\gamma$, so that $E_{3}\tau ^{\prime }_3/m_3$ is the decay 
time of particle $3$ with moment $p_{3}$ 
\begin{equation}
\frac{dW_{3\rightarrow 1+2}}{dVdt}=\frac{g_3}{\left( 2\pi \right) ^{3}}\int
d^{3}p_{3}f_{b,f}(\Upsilon _{3},p_{3})\frac{m_{3}}{E_{3}}\frac{1}{\tau^{\prime }_3}.  \label{Decay1}
\end{equation}%
 
Comparing Eq.(\ref{Decay1}) with   Eq.(\ref{pp}), we conclude
that in medium, at finite temperature $T$,  the 
decay rate  $\tau_3^{\prime }$ of particle $C_3$ in the rest frame of  the heat bath 
is given by
\begin{widetext}
\begin{align}
\frac{1}{\tau ^{\prime }_3}& =\frac{1}{2m_{3}}\frac{1}{1+I} \int \frac{
d^{3}p_{1}}{2E_{1}\left( 2\pi \right) ^{3}}\int \frac{d^{2}p_{2}}{2E_{2}\left(2\pi \right) ^{3}}
 \left( 2\pi \right) ^{4}\delta ^{4}\left(p_{1}+p_{2}-p_{3}\right) \notag \\[0.2cm]
&
\times \frac{1}{g_3}\sum_{spin}\left\vert \langle p_{1}p_{2}\left\vert
M\right\vert p_{3}\rangle \right\vert ^{2} f_{b,f}(\Upsilon _{1},p_{1})f_{b,f}(\Upsilon _{2},p_{2})
\Upsilon_{1}^{-1}\Upsilon _{2}^{-1}\exp (u\cdot p_{3}/T),  \label{tau23}
\end{align}%
\end{widetext}
which is a Lorentz invariant form, but $ u\cdot p_{3}\to E_3$, the energy of the
particle $3$ in the rest frame of the thermal bath.

Using the in-vacuum particle $C_3$ rest-frame decay time, Eq.(\ref {VacTau}), 
we find that  Eq.(\ref{tau23}) takes the form:
\begin{equation}
\frac{1}{\tau ^{\prime }_3}=\frac{1}{\tau _{0}}\frac{e^{E_{3}/T}}{2}\Phi (p_{3}).
\label{tau-tau0}
\end{equation}%
 The function $\Phi (p_{3})$ is:
\begin{equation}
\Phi \left( p_{3}\right) =\int_{-1}^{1}d\zeta
\frac{\Upsilon_{1}^{-1}}{\Upsilon _{1}^{-1}e^{\left( a_{1}-b\zeta \right) }\pm 1}
\frac{\Upsilon_{2}^{-1}}{\Upsilon _{2}^{-1}e^{\left( a_{2}+b\zeta \right) }\pm 1}.
\label{phif}
\end{equation}
with
\begin{eqnarray}
&&a_{1}=\frac{E_{1}E_{3}}{m_{3}T}, \quad
a_{2} =\frac{E_{2}E_{3}}{m_{3}T}, \quad
b =\frac{pp_{3}}{m_{3}T}\quad {\rm and}\nonumber\\
&&\zeta =\cos \theta =\cos (\vec{p}_{2}\wedge \vec{p}_{1}).
\end{eqnarray}
With this the  particle $C_3$ decay rate per unit volume in a thermally
equilibrated system is given by
\begin{equation}
\frac{dW_{3\rightarrow 1+2}}{dVdt}=\frac{g_3}{\left(2\pi^{2}\right) }
\frac{m_{3}}{\tau _{0}}\int_{0}^{\infty }\frac{p_{3}^{2}dp_{3}}{E_{3}}
\frac{e^{E_{3}/T}}{\Upsilon_{3}^{-1}e^{E_{3}/T}\pm 1}\Phi (p_{3}),  \label{Decay1-final}
\end{equation}

We were able to evaluate the integral $\Phi(p_3)$  analytically in the
absence of particle-antiparticle asymmetry (absence of chemical potentials),
\begin{eqnarray}
 \Phi(p_3)&=&\frac{1}{b(e^{a_1+a_2} \pm \Upsilon_1\Upsilon_2)}\times\nonumber\\
 &&\ln\frac{\left(\Upsilon_2e^{-a_2} \pm e^b\right)
\left(e^{a_1} \pm \Upsilon_1e^{-b}\right)}{\left(\Upsilon_2e^{-a_2}\pm e^{-b}\right)\left(e^{a_1} \pm \Upsilon_1e^b\right)}. \label{phia1a2b}
\end{eqnarray}
and in the non-relativistic limit ($m_{3}\gg T, p_3$), this quantity
tends to
\begin{equation}
\Phi (p_3\to 0) = 2\frac{\Upsilon _{1}^{-1}\Upsilon_{2}^{-1}}
 {(\Upsilon_{1}^{-1}e^{E_{1}/T}\pm 1)(\Upsilon _{2}^{-1}e^{E_{2}/T}\pm 1)}.
\label{philim}
\end{equation}

\subsection{Decay and production rates in the Boltzmann limit}
A useful check of the more complex quantum decay case is the   Boltzmann limit.
We can then omit unity in the distribution Eq.(\ref{f}). This is possible when
\begin{equation}
\Upsilon _{i}^{-1}e^{u\cdot p_{i}/T}\gg 1,  \label{Boltzmann}
\end{equation}%
that is, when $\Upsilon _{i}\ll 1$ or $T\ll m_{3}/2$. The condition $T\ll m_{3}/2$
comes from the fact that the minimal energy of lighter particles is
$m_{3}/2$ in the particle $3$ rest frame. In this limit the decay time in the
particle $3$ rest frame from Eq.(\ref{tau23}) $\tau ^{\prime }\rightarrow\tau _{0}$
 so that from Eq.(\ref{taui}) we have, for the average decay rate
$\tau $ in the reference frame (the rest frame of the bath),
\begin{eqnarray}
\tau_3^\prime  &\approx &\tau _{0}
\frac{\int_{0}^{\infty }p^{2}dp \,e^{{E_{3}}/{T}}}
        {\int_{0}^{\infty }p^{2}dp \, e^{{E_{3}}/{T}}{m_{3}}/E_{3}} \\[0.3cm]
&=&\tau _{0}\frac{K_{2}(m_{1}/T)}{K_{1}(m_{1}/T)}.  \label{taubl}
\end{eqnarray}

Equation (\ref{Decay1}) shows that the average decay time $\tau_3^\prime $ 
in the laboratory frame
is proportional to the  (inverse) average of the Lorentz factor of particle
$C_3$. We address this effect next    in a quantitative manner; 
 the ratio of $\tau_3^\prime  $ to $\tau _{0}$ is shown
in Fig.{\ref{taurho}} as the dotted line. For $T\ll m_{3}$ this ratio goes to
unity because the Lorentz factor becomes $1$. For large $T $, the
rate increases because of the higher average energy of particle $C_3$,
that is, the increasing   average Lorentz factor $\gamma$. Therefore, for the 
low density  classical limit with  $\Upsilon _{i}\ll 1$  the average
particle life time increases with $T$ owing to relativistic effects.
However a different result can arise for a dense quantum medium.

\section{Examples}\label{4}
\subsection{Hadrons in heavy-ion collisions}
\subsubsection{Production of $\protect\rho $ mesons via the $\protect\rho %
\leftrightarrow \protect\pi \protect\pi $ process}

First, we consider an example of $\rho$-meson thermal decay and production 
in a thermal and chemically equilibrated pion bath:
\begin{eqnarray}
&&\rho ^{0}\leftrightarrow \pi ^{+}+\pi ^{-},  \label{ropi++} \\
&&\rho ^{\pm }\leftrightarrow \pi ^{\pm }+\pi ^{0}.  \label{ropi+0}
\end{eqnarray}%
In this example all particles are bosons and we put $m_{1}=m_{2}$  
for simplicity, which is not quite exact  for reaction (\ref{ropi+0}).
In integral (\ref{phif}) we have $E_{1}=E_{2}=m_{\rho }/2$ in the $\rho$ rest
frame. The integrand in $\Phi(p)$ is a symmetric function. Then we can write
\begin{equation}
\Phi \left( p_{\rho }\right) =2\int_{0}^{1}d\zeta
\frac{\Upsilon _{\pi }^{-2}}{\Upsilon _{\pi }^{-1}e^{\left( a-b\zeta \right) }-1}
\frac{1}{\Upsilon_{\pi }^{-1}e^{\left( a+b\zeta \right) }-1}.  \label{phif2}
\end{equation}
where
\begin{equation}
a =\frac{\sqrt{m_{\rho }^{2}+p_{\rho }^{2}}}{2T};\quad
b =\frac{\sqrt{1-4m_{\pi }^{2}/m_{\rho }^{2}}p_{\rho }}{2T}.
\end{equation}
The integral, Eq. (\ref{phif2}), can be evaluated in this case as
\begin{eqnarray}
\Phi (p_{\rho })&=&
\frac{2\Upsilon _{\pi }^{-2}}{b(\Upsilon _{\pi}^{-2}e^{2a}-1)}    \nonumber\\
&&\hspace{-0.8cm}\times\left(b+\ln \left( 1+
\frac{ \Upsilon _{\pi}\left(e^{(b-a)}-e^{-(a+b)}\right) }
  {\left(1-\Upsilon_{\pi }e^{b-a}\right) }\right) \right).  \label{phiab}
\end{eqnarray}%
Then we substitute $\Phi$ into Eq.(\ref{Decay1}) and using Eq.(\ref{pdr})
we can calculate $\rho $ decay and production rates. To calculate $\tau_3\to \tau$
we use definition (\ref{taui}).

\begin{figure}[tbp]
\centering \includegraphics[width=8.6cm,height=8.5cm]{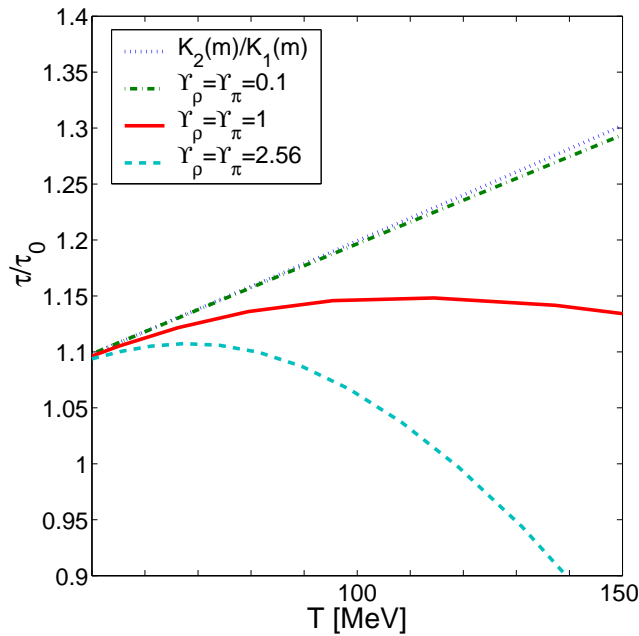}
\caption{\small (Color online) The ratio $\protect\tau  /\protect\tau _{0}$
as a function of temperature $T$   in the
reaction $\rho\leftrightarrow \pi\pi$. 
The  [dotted (blue) line] is
for  the Boltzmann limit showing only time dilation. 
Near this limit  [dash-dotted (green) line]
dilute system   $\Upsilon_{\rho}=\Upsilon_{\pi}=0.1$. 
Solid (red) line and dashed (turquoise) lines represent
$\protect\Upsilon_{\rho} = \Upsilon_{\pi}=1$ 
and   $\Upsilon_{\rho} =\Upsilon_{\pi}= 2.56$, respectively. }\label{taurho}
\end{figure}

In Fig. {\ref{taurho}} we present the $\rho $ decay time 
in the laboratory frame normalized by its decay time in the rest frame in a vacuum  as a
function of temperature $T$ for $\Upsilon _{\rho }=\Upsilon _{\pi }=1$,
[solid (red)  line], $\Upsilon _{\rho }=\Upsilon _{\pi }=2.56$ 
(dashed line), $\Upsilon _{\rho }=\Upsilon _{\pi }=0.1$ (dash-dot 
line); the dotted line shows the Boltzmann limit, Eq.(\ref{taubl}). We consider the range of
temperatures between $50$ and $150$ MeV, which includes the QGP
hadronization temperature ($\approx$ 140 -165 MeV). 

We show the case $\Upsilon _{\rho }=\Upsilon _{\pi}=0.1$ (dot-dashed line) 
to check the transition to the Boltzmann limit. We can see that for this case, the
result is close to the Boltzmann approximation for our range of $T$, as 
expected. In the case of  chemical equilibrium, $\Upsilon _{\rho }=\Upsilon _{\pi }=1$,
the solid line in Fig.~{\ref{taurho}} shows a relatively small, 10-15\% increase in lifie span. 
We finally consider a supersaturated pion state, 
$\Upsilon _{\rho }=\Upsilon _{\pi }=2.56$, 
which can arise after supercooled QGP 
hadronization near  $T=140$ MeV~\cite{Kuznetsova:2006bh}.
For small $T\ll m_{\rho }/2$, the ratio $\tau /\tau _{0}$ is near the Boltzmann
limit, close to unity, because for such a small $T$, when the Boltzmann limit is applied,
decay time $\tau $ does not depend on $\Upsilon$. When $T$ increases quantum the
effects dominate and $\tau $ decreases with increasing $T$. In general, 
the larger $\Upsilon$ is, the more rapidly $\tau$ decreases with temperature. 

Here we do not consider in depth the  $\rho$-meson density evolution 
in heavy-ions collisions, because without doubt, our limited 
system (just a few hadron states) is not sufficiently realistic to 
capture the physics  of the $\rho$ in dense matter. Moreover, the
$\rho$ yield is  a probe of the hadron density temporal evolution and
thus still more difficult to describe precisely. This can be seen as 
follows: In the Boltzmann low-density limit, the chemical 
equilibration time is  $\tau_\rho \approx$ 1.7 fm. 
Our result shows that the pion high density quantum medium effects 
causes an increase in $\rho$ width, that 
is a decrease in equilibration time to $\tau_\rho \approx$ 1.25 fm, 
accelerating $\rho$-meson chemical equilibration near the hadronization 
temperature. 
The kinetic-phase time scale in heavy-ion collisions, when hadrons 
interact is near 2-3 fm~\cite{Kuznetsova:2008zr}.
This means that the  $\rho$-meson
chemical evolution  is dependent on the ambient hadron density
and thus is intricately connected  with the dynamics of fireball expansion.
 
\subsubsection{$\phi$ meson evolution in heavy ion  collisions}
We consider here $\phi$-meson yield evolution in a thermal 
hadronic gas after QGP hadronization 
formed in heavy-ion collisions.  
The temperature of QGP hadronization can be within the range 140-180 MeV. 
After hadronization, individual hadrons can continue to rescatter into resonances in what
we call the kinetic  evolution phase or thermal hadronic gas. This scattering effect does not
materially change the final stable particle yields, but it affects the yields of
resonances observed by the invariant mass method. The temperature of kinetic-phase freeze-out 
is expected to be near 100 MeV. After kinetic freeze-out, hadrons
expand without interactions, via decay only. 

The $\phi$-meson has, on the hadron reaction scale a
relatively small width $\Gamma_\phi \approx 4.26$ MeV,
that is $\tau_\phi \approx$ 46 fm, which  is much 
longer than the duration of the kinetic phase. 
About 83$\%$ of $\phi$-mesons decay into $K+K$. 
Therefore we consider here $\phi$ evolution in the
reaction:
\begin{equation}
K+K \leftrightarrow \phi. \label{phiKK}
\end{equation}
We do not consider here the decay channel $\phi \rightarrow \rho+\pi$, 
which is about 15$\%$ and  can influence our result at this  level. 
Moreover, the $\phi$ inelastic scattering in two-to-two particle 
reactions also has a noticeable influence on $\phi$  yield, 
about $15\%$ suppresion~\cite{AlvarezRuso:2002wx}. 
In~\cite{AlvarezRuso:2002wx}  only the $\phi$ decay 
was included, without the reverse reaction, 
assuming an initial equilibrium yield at hadronization. 
We show here how the reverse reaction and nonequilibrium 
hadronization conditions can influence the resulting 
$\phi$ yield. The effect from the full one-to-two reactions, \req{phiKK}, can be added 
to that from two-to-two particles reactions.

We did a similar study previously for baryon resonances $\Delta(1232)$ and $\Sigma(1385)$~\cite{previously:2008zr}. We found that in the case of 
initial non-equilibrium yield, when we have an overabundance of stable particles
 at hadronization, the resonance production can be greater than the resonance decay.
Decay becomes dominant when the temperature drops with expansion, and the lighter mass 
decay product states becomes statistically preferable. The final resonance yield depends 
on the study of the balance between these two effects.  

The $\phi$-meson width is smaller than the width of these baryon resonances, and its 
yield change during the posthadronization kinetic phase is expected to be smaller. 
However, the rather low threshold energy, $m_{\phi}-m_K-m_K \approx 30$ MeV, could
mean  that $\phi$ production is dominant over a longer period of time than in the aforementioned case of baryon resonance. The purpose of this short study is to determine 
how much the yield of $\phi$ can change during the kinetic phase owing to kaons fusion compared to 
its yield at hadronization.   We do not study
 how relativistic and quantum effects influence the reaction, Eq. (\ref{phiKK}),
relaxation time, because for the range of temperature considered these effects are small.

Considering that the mass of all particles involved is greater than the temperature 
it is possible to  use the  Boltzmann distribution for $\phi$ and $K$:
\begin{eqnarray}
\frac {N_{\phi} }V&=&\Upsilon_{\phi}\frac{T^3}{2\pi^2}g_{\phi}x_{\phi}^2K_2(x_{\phi}), \\
\frac {N_{K} }V&=&\Upsilon_{K}\frac{T^3}{2\pi^2}g_{K}x_{K}^2K_2(x_{K}),
\end{eqnarray}
where $x_{i}=m_{i}/T$ and $K_2(x)$ is the Bessel function. We proceed as in Ref.~\cite{Kuznetsova:2008zr},
using Eq.(\ref{uppiu}). 
 
Initial conditions in the kinetic phase are defined 
by conditions at QGP hadronization.
We introduce  the  initial hadron yields in a framework of a rapid
QGP hadronization with all hadrons produced with
yields governed by the entropy and strangeness
content of QGP by quark recombination. In this model
the yields  of mesons and baryons are controlled by
the  constituent  quark fugacity $\gamma_q$:
\begin{equation}
\Upsilon^0_{K}=\gamma_q\gamma_s; \qquad \label{upinpi}
\Upsilon^0_{\phi}=\gamma_s^{2}. 
\end{equation}
Thus for $\gamma_q>1$ we have the condition $\Upsilon_{\phi}<\Upsilon_{K} \Upsilon_{K}$.
At first, the reaction goes toward $\phi$ production until the $\phi$ density reaches the 
equilibrium point when the right hand side of Eq.(\ref{uppiu})
is 0. If the $\phi$ density has enough time to reach this point, 
it begins to decrease again because the temperature decreases owing to expansion.

For each entropy content of the QGP fireball, the corresponding fixed 
background value of $\gamma_q$ can
be found once the hadronization temperature is known~\cite{Kuznetsova:2006bh}.  
For $T=140$ MeV pions form a  nearly fully  degenerate Bose gas with 
$\gamma_q\simeq 1.6$.
In the following discussion, besides this  initial condition, 
we also consider the value pairs
$T=160{\rm\, MeV},\,\gamma_q=1.27$
and $T=180$ MeV with $\gamma_q=1$. The value of $\gamma_s \ge 1 $ 
plays no significant role as, in the reaction considered, \req{phiKK}
the number of strange quarks present is the same. 

\begin{figure}
\centering
\includegraphics[width=8.3 cm, height=8.3 cm]{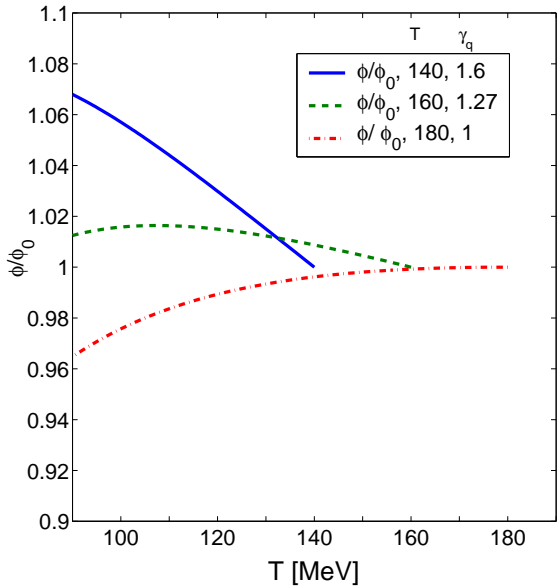}
\caption{\small{(Color online) The yield ratio ${\phi}/{\phi}^0$ for 
hadronization temperatures $T_0=140$ MeV [solid (blue) line], 
$T_0=160$ MeV [dashed (green) line]
and $T_0=180$ MeV [dash-dotted (red) line] as functions of ambient
temperature $T$.
}}
\label{phifig}
\end{figure}

In Fig. \ref{phifig} we present results for the ratios
${\phi}/{\phi}_0$ at different hadronization temperatures as
functions of temperature $T$, beginning from the presumed initial hadronization
temperature $T_0$ through  $T_ {\rm min}= 90$ MeV.
$\phi^0$ is the initial yields obtained at each hadronization
temperature.  For hadronization temperatures 
$T_0 < 180$ MeV ($\gamma_q=1.6$), we initially have $\Upsilon_\phi < \Upsilon_K\Upsilon_K$.
In these cases the master equation leads to an initial increase 
in the yield of resonances. In the case $T_0=140$ MeV, when the effect is largest, this
increase in $\phi$ yield continues over the full range of temperature considered. 
However, the effect is relatively small, about $7\%$, owing to the small $\phi$ width.
For hadronization temperature $T=160$ MeV, when $\gamma_q = 1.27$  is smaller, 
the increase in yield is smaller, and at $T=105$ MeV the $\phi$ yield begins 
to decrease slowly owing to the dynamics of the expansion.

We note that for $T \ge 180$ MeV there is always 
a slow depletion of the $\phi$ resonance yield. This result implies that the observed
yield of $\phi$ has a systematic +7\%/-4\% uncertainty due to kaon 
rescattering in the medium. 
For comparison in~\cite{AlvarezRuso:2002wx} the effect from $\phi$ decay
only for equilibrium yield at hadronization ($\gamma_q=\gamma_s=1$) was
determined to be   -$7.5\%$ and the  effect from 2-to-2 particles 
reactions was -15$\%$. Therefore $\phi$ production in kaon fusion 
for nonequilibrium hadronization conditions may have an enhancement  effect of about 15\%
  on the final $\phi$ yield, compared 
to the scenario where the $\phi$ can only decay 
after in-equilibrium hadronization formation.

We cannot compare with experimental results considering 
only the kaon fusion reaction, as it was argued in Ref.~\cite{AlvarezRuso:2002wx}
that certain  two-to-two reactions and possibly other processes 
can influence the yield. We note only that kaon fusion can 
add to the observed $\phi$ yield~\cite{Abelev:2008fd}.

\subsection{Freeze-out processes in the early Universe}
\subsubsection{ $\protect\pi ^{0}$  at $T\ll m_\pi$}

As mentioned in Sec.I, it is interesting to examine the mean life
time of $\pi^{0}$ in the end of the hadronic-gas stage of the Universe where
the temperature drops to a mega-electron-volt level in the low teens. Then the reaction%
\begin{equation*}
\pi ^{0} \leftrightarrow \gamma +\gamma
\end{equation*}%
determines the abundance of $\pi ^{0}$.

The difference from the previous example is that the photons are
massless and they are always in chemical equilibrium in the early Universe
($\Upsilon_{1}=\Upsilon_2=1$). Then we can rewrite function
(\ref{phia1a2b}) as
\begin{equation}
\Phi (p_{\pi^0})=\frac{2}{b(e^{2a}-1)} \left( b+\ln \left(1+\frac{ 
e^{(b-a)}-e^{-(a+b)} }{ 1-e^{b-a}  }%
\right) \right).  \label{phiab1}
\end{equation}%
with
\begin{equation}
a =\frac{\sqrt{m_{\pi^0 }^{2}+p_{\pi^0 }^{2}}}{2T}; \quad
b=\frac{p_{\pi^0}}{2T}.
\end{equation}
Again we use Eq.(\ref{Decay1}) and (\ref{pdr})
we can calculate $\pi_0$ decay and production rates. To calculate $\tau$
for $\pi^0$ decay in matter  we use definition (\ref{taui}).

\begin{figure}[tbp]
\centering \includegraphics[width=8.6cm,height = 8.5cm]{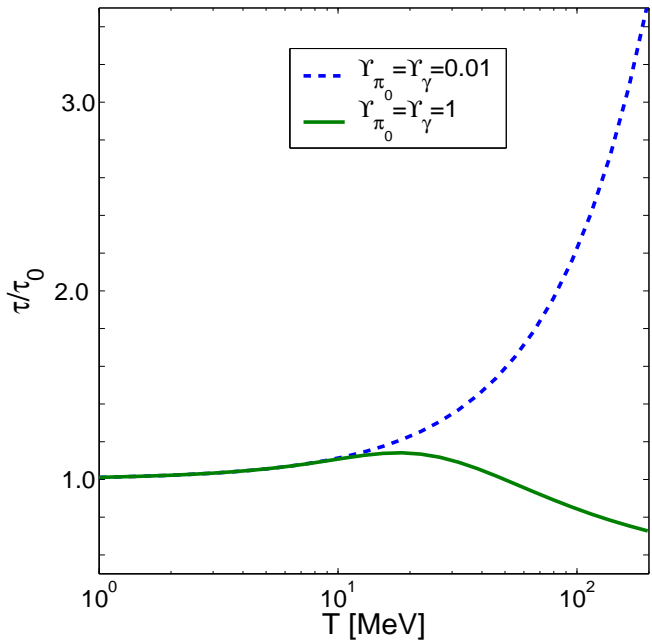}
\caption{{\protect\small {(Color online) The ratio $\protect\tau /\protect\tau
_{0}$ for $\pi^0$ decay/production as a function of temperature $T$.
Dashed (blue) line is for a dilute system  
$\protect\Upsilon_{\pi^0}=\Upsilon_{\gamma}=0.01$ (Boltzmann limit); 
solid (green) line is for a thermal chemically equilibrated system 
$\Upsilon_{\pi^0}=1$.
}}} \label{taupi0}
\end{figure}

In figure \ref{taupi0} we show ratio of $\pi^0$ decay time $\tau_3\to \tau$ in the
presence of thermal particles to the decay time in vacuum in $\pi_0$
rest frame: $\tau_{\pi^0}/\tau^0_{\pi^0}$. In this figure a wider
range of temperature is shown $1-200$ MeV. For
$\Upsilon_{\pi^0}=1$ the ratio $\tau_{\pi^0}/\tau^0_{\pi^0}$ the
temperature dependence is similar to that for $\rho$ decay,
considered in previous chapter. It increases at first due to 
relativistic time dilution effects. Then, after $T\approx
20$ MeV, $\tau$ goes down slowly with temperature, when the quantum in-medium
effect becomes important. Quantum in-medium effects arise here 
 mostly from photons. They compensate relativistic  Lorentz factor
effect when $T$ is about $m_{\pi}$; compare the lines in Fig.~\ref{taupi0} for $\Upsilon_{\pi}=\Upsilon_{\gamma}=0.01$ 
[dashed (blue)] and  $\Upsilon_{\pi}=\Upsilon_{\gamma}=1$ [solid (green)]. Note that 
when the yield of pions is small, that is,  only $\Upsilon_{\pi}$ is small, 
the  result is almost the same as in chemical equilibrium, $\Upsilon_{\pi}=\Upsilon_{\gamma}=1$.

As long as the $\pi_0$ reaction relaxation time $\tau$ is much shorter compared 
to the Hubble expansion time  $T/\dot T=1/H$, there is  chemical equilibrium 
in the Universe with $\Upsilon_{\pi^0}=1$. 
Freeze-out from chemical equilibrium arises when condition Eq.(\ref{frcon}) is satisfied.
Because  $\tau \approx \tau_0 = 8.4\,10^{-17}$s the condition Eq. (\ref{frcon}) is always 
satisfied where $\pi^0$ can exist. Only at unrealistically large temperatures this 
condition can be violated.
 
Therefore we conclude that for the temperature range of interest, 
between few MeV and 180 MeV, the $\pi^0$ are in chemical equilibrium 
with photons because of their fast reaction  rate. Note that
weak  interaction process such as neutron decay $n \rightarrow p + e^- + \nu_e$
is 20 orders of magnitude slower, and the Universe expansion rate can dominate 
the neutron decay rate e.g. at $T > 0.1$ MeV, before having a good chance to decay,
neutrons are thus available to enter nuclear reactions. 

The pion,  and muons equilibrium density is  large 
until temperatures near a few MeV, and as a result, 
they participate in reactions with each other, 
nucleons and other particles; for example see the next sections.
The importance of this  realization of pion chemical equilibrium in the Universe
is that it implies that all hadron species, driven by pions also remain 
in chemical equilibrium.
Their  abundances can thus be computed using the chemical 
equilibrium hypothesis, as done  in Ref.\cite{FromerthRafelski}.

\subsubsection{$\pi^{\pm}$, $\mu^{\pm}$, and $\nu,\,\bar\nu$  equilibration/freeze-out}

In the laboratory the dominant   $\pi^{\pm}$  production reaction 
is the pion charge exchange reaction, Eq. (\ref{pipm}), which we 
considered in Ref.~\cite{Kuznetsova:2008jt}. These reactions also 
can take place in  the  early Universe. However, 
given the much slower  evolution of the  early Universe, 
we also now encounter reactions involving 
neutrinos, Eq. (\ref{pimunu}); the related in-vacuum,  
weak-decay lifespan of the  $\pi^{\pm}$  is  $\tau_0=2.60\times10^{-8}$ s. 

In Fig.~\ref{pipmmunu} we show the relaxation times in units of $\tau_0$ for $\pi^{\pm}$  
equilibration, Eq. (\ref{taui}),  in reaction (\ref{pimunu}), as functions of temperature:
near $T\simeq 160$ MeV the lifespan is enhanced by a factor of 3 
for thermal equilibrium densities with $\Upsilon$s =1  
[solid (blue) line in figure~\ref{pipmmunu}] mostly owing to Pauli blocking of the 
decay products.  The time dilation owing
to thermal motion, which also prolongs the life span, 
has a smaller effect, visible in the Boltzmann limit, 
which we study for a dilute system with $\Upsilon$s=0.01, [dashed (green) line]. 

\begin{figure}
\centering \includegraphics[width=8.6cm,height = 8.5cm]{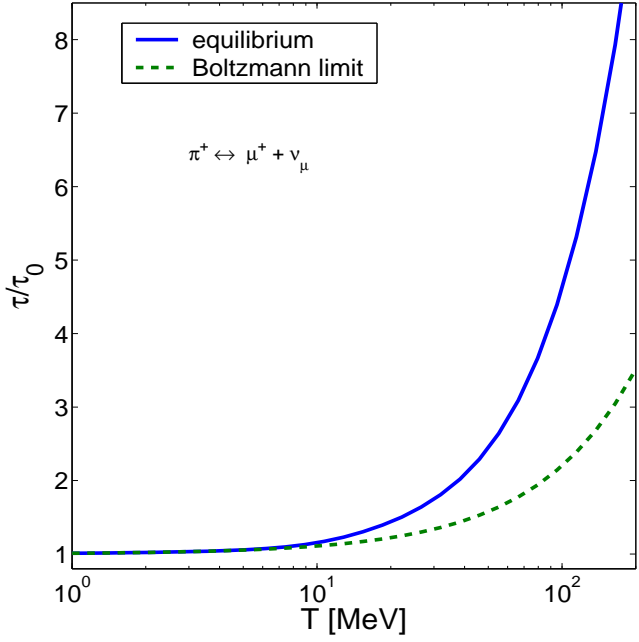}
\caption{{\protect\small {(Color online) $\pi^{\pm}$  relaxation time 
as a functions of $T$ in reaction Eq.(\ref{pimunu}),
in thermal equilibrium [solid (blue) line]  and in the Boltzmann limit,
 obtained for $\Upsilon$ =0.01  [dashed (green) line].}}} \label{pipmmunu}
\end{figure}

Interestingly, as we next show,
the process, Eq.(\ref{pimunu}), is the fastest mechanism of neutrino equilibration
in a wide range of temperatures relevant here, $T>7$ MeV, but the  $\nu_\mu$-freeze-out condition
is at a lower $T$ and seems to be controlled by the reaction~\cite{Freese:1982ci}
\begin{equation}
e^++e^-\leftrightarrow \nu_{e,\mu} +\bar\nu_{e,\mu}  \label{eenunu}
\end{equation}
which we also consider now. The neutrino oscillation effect assures that 
all neutrinos  remain in equilibrium as long as one is strongly coupled to 
the system. 

\begin{figure}
\centering \includegraphics[width=8.6cm,height = 8.5cm]{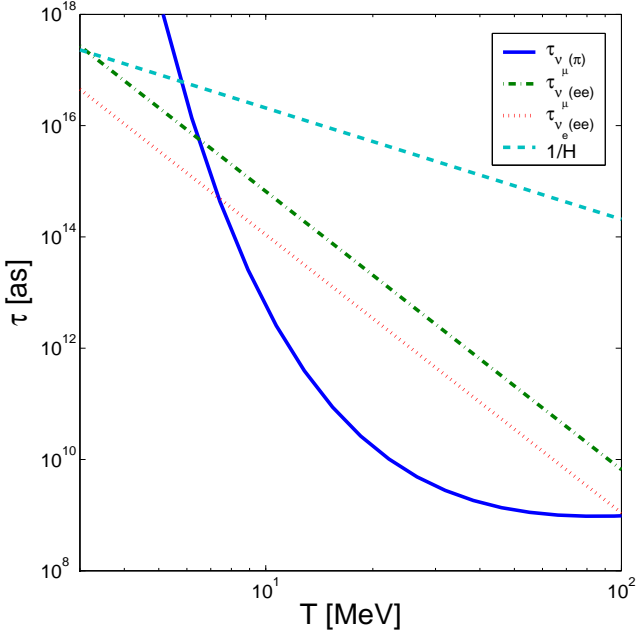}
\caption{{\protect\small {(Color online) Relaxation time for neutrino $\nu_{\mu}$ 
equilibration as function of temperature compared to the
Universe expansion time $1/H$ [dashed (turquoise) line]. 
Solid (blue) line   for reaction Eq.(\ref{pimunu}) with equilibrium densities ($\Upsilon$ =1);
dash-dotted (green) line , and dotted (red) line   are for reaction  (\ref{eenunu}) for muon and
 electron neutrino, respectively.}}} \label{tauneut}
\end{figure}

In Fig.~\ref{tauneut} we show muon-neutrino equilibration 
time in reaction Eq.(\ref{pimunu}) (solid line, blue);
recall that to obtain this relaxation time from the results shown in Fig. \ref{pipmmunu},
we  need to replace the  $\pi$ density  in the nominator of Eq.(\ref{taui}) 
by the density of $\nu$. This relaxation time intersects the Universe
expansion rate at $T \approx 5.5$ MeV. Freese et al~\cite{Freese:1982ci} obtain  the 
relaxation time as a function of $T$ assuming the neutrino chemical potential $\mu_{\nu}\ll T$
in reaction \req{eenunu}:
\begin{eqnarray}
\tau_{\nu_{\mu}(ee)} = (0.1G_F T^5)^{-1},\quad
\tau_{\nu_{e}(ee)} = (0.6G_F T^5)^{-1},
\end{eqnarray}
where $G_F=1.1664\,10^{-5}$ GeV$^{-2}$ is the Fermi constant. These two results
are shown in Fig. \ref{tauneut}. We see that the muon-neutrino freeze-out
temperature according to reaction (\ref{pimunu}) is slightly higher than that according 
to reaction (\ref{eenunu}). The temperature of the neutrino decoupling in reaction 
\req{eenunu} is $T_d \cong 3.5$ MeV for $\nu_{\mu}$ and $T_d \cong 2.0$ MeV for $\nu_{e}$.

For a wide range of temperatures, to as low as 7 MeV neutrino
chemical equilibration by reaction (\ref{pimunu})
is dominant. This example shows that reactions with chemically equilibrated 
pions and muons can have an influence on other, even lighter particle 
evolution for temperatures $T<<m$. 

Muons can be equilibrated by reaction (\ref{muprod})  and by the $1 \leftrightarrow 3$
reaction [Eq.(\ref{munu})].  We do not consider this type of reaction in detail  here. 
For low temperatures, $T\ll m_{\mu}$, when relativistic and medium effects are small,
we assume that the muon decay time and reverse reaction relaxation time
are nearly the muon lifespan in vacuum, $\tau_0=2.20\, 10^{-6}$s. 

\begin{figure}
\centering \includegraphics[width=8.6cm,height = 8.5cm]{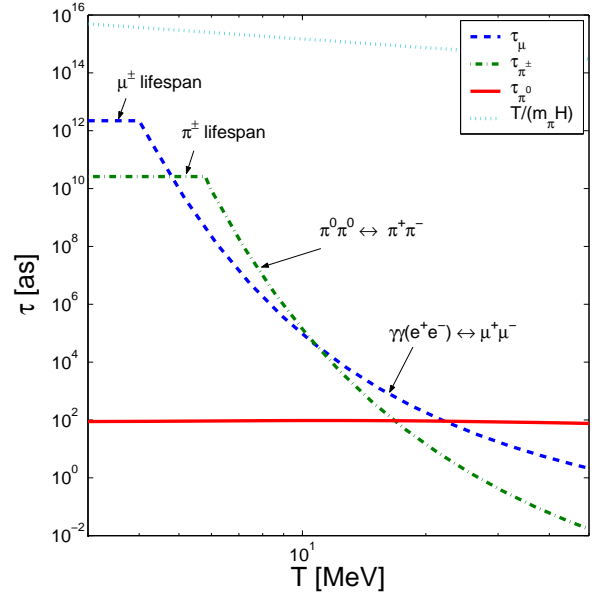}
\caption{{\protect\small {(Color online) Equilibration times as functions of temperature,
for $\pi^0$ (solid line, red), $\pi^{\pm}$ [dash-dot (green) line] , 
$\mu^{\pm}$ (dashed line, blue) and 
$\tau_{T} \approx T/(Hm_{\pi})$ [dotted (turquoise) line].}}} \label{taupih}
\end{figure}

In Fig.~\ref{taupih} we show relaxation times for dominant 
reactions for pion and muon equilibration.  For $\pi^\pm$
reaction, Eq.(\ref{pimunu}), becomes dominant 
over reaction (\ref{pipm}) at $T\approx  6$ MeV. 
For the $\mu^\pm$   reaction (\ref{munu}) becomes dominant 
at $T\approx 4$ MeV. Therefore at these low temperatures, 
relaxation times for  $\mu^\pm$  and $\pi^\pm$ equilibration
 becomes constant and far below the Universe expansion rate 
and $\tau_{T}$ [dotted (turquoise) line]. We conclude that $\mu^\pm$ and $\pi^\pm$
stay in chemical equilibrium. This does not mean that they play an 
important role in the global physics of the early Universe,
because just at these temperatures muon and pion densities begin 
to drop rapidly and  soon their yield is negligibly small, 
far below the nucleon  density in the Universe.

\section{Conclusions}
We have presented detail of the kinetic master equation for a
process involving the formation of an unstable particle through reaction
(\ref{123}) in a relativistically covariant fashion. Assuming that
all particles in the process are in thermal equilibrium, we
calculated the thermal averaged decay and formation rates of the
unstable particle. Using the time reversal
symmetry of quantum processes, we have shown 
that the time evolution of the density of an 
unstable particle is given by (\ref{fe}).
Therefore in chemical equilibrium the particle fugacities are connected by  
(\ref{equilcon}). We have explicitly derived the thermal decay rate
of an unstable particle, obtaining  \req{Decay1-final}.

The general properties of
the thermal particle decay/production kinetics have led us
to consider the relaxation time defined by \req{taui}, which results
in a greatly simplified kinetic equation \req{uppiu}. The 
medium modification of reaction rates we encountered are all caused by 
final-state quantum effects, Bose enhancement, and/or Fermi blocking,
absent in the Boltzmann limit. Moreover, we note the presence of 
kinematic effects, in that all lifespans of particles are time dilated
owing to their motion with respect to the thermal bath rest frame. 

In the present formalism, we assumed that the decay width of an
unstable particle is much smaller than the temperature $T$. This
approximation is safe   in the examples we have discussed above, except 
perhaps the case  of $\rho$ decay, where some corrections may be needed.
For the formation of heavy resonances, whose decay width
becomes appreciable compared to the temperature $T$, we may need to
include the finite-width effect on the mass of an unstable particle
in the thermal distribution. Such effects on the statistical partition
function and the equation of state of a system have been studied based
on the virial expansion method \cite{Width}. The correction for a
kinetic equation in such a case has also been studied \cite{Width2}.

We have presented several  examples,  
$\rho \leftrightarrow \pi +\pi$,
$\phi  \leftrightarrow {\rm K}+ \overline {\rm K}$,
$\pi ^{0}\leftrightarrow \gamma +\gamma$,  and 
$\pi^{\pm} \leftrightarrow \mu^{\pm} + \nu_{\mu}(\bar\nu_{\mu})$,
and explored the physics cases of hot hadron matter created 
in laboratory heavy ion reactions,
and the early Universe from the condition of hadronization down to the temperature 
of several mega-electron volts. The two first processes can take
place in both circumstances.   The third
process is important to the understanding of how the 
hadronic fraction evolves  with the expansion of 
the Universe. The last process we considered appears to
be the dominant mechanism of neutrino equilibration over the
entire temperature range, except close to neutrino 
freeze-out, a result that requires further refinement allowing
for finite chemical potentials.  This example also shows that  
heavy ($m >> T$) chemically equilibrated particles can be important 
in the evolution of other particles including lighter more dense particles
yields at relatively low temperatures.

The equilibration-relaxation time 
for $\pi^0$ decay remains close (within 25\%)  to the relaxation
time in vacuum for a large temperature range. This occurs because the 
relativistic effect (Lorentz factor) is compensated by the quantum 
medium effect. This  time is short compared to the Universe 
expansion time for all temperatures of interest here, below the 
QGP hadronization temperature,  when $\pi^0$ hadrons are created. Therefore 
$\pi^0$ always stays in chemical equilibrium with radiation 
for the temperature range of interest.

As long as $\pi^0$ is abundant it  can participate in reactions
 with other hadrons and influence the dynamics of the Universe evolution.
Here we also considered $\pi^{\pm}$ evolution, in the fourth reaction given 
and their interaction with $\pi^{0}$.  We showed 
that pions  and muons (mesons) stay in chemical equilibrium throughout the evolution of
Universe, despite their large mass. They can be involved 
in reactions with nucleons, a topic we postpone to a
future study, down to temperatures where the meson density 
drops well below the nucleon density. The contribution of mesons  
disappears from the entropy and the degeneracy $g$ only at the relatively 
low $T \approx 10$ MeV (see Fig.~\ref{degen}). 

Our study of the  $\phi$ evolution in thermal hadron medium
after QGP hadronization in heavy ions collisions,
suggests a possible  slight modification of the observed $\phi$  yield:
compared to initial production, an increase in hadronization 
at   $T=140$ MeV, $\gamma_q=1.6$, at the level of  about 6\%-7\%,
or a suppression of about 4\%  for hadronization at  $T=180$ MeV, $\gamma_q=1$.

To conclude, we have presented here  the process of decay and 
and re-creation of unstable particles, and studied
special cases of relevance to heavy-ion collisions and the early 
Universe. Our results indicate that the early Universe was in
chemical equilibrium throughout its evolution and that the first
freeze-out occurs when neutrinos decouple. \\

\acknowledgments
We thank H.Th. Elze, M.J. Fromerth,  T. Kodama, J. Letessier, M. Makler,
and R.L. Thews, for valuable discussions regarding hadron phase
chemical equilibration in the early Universe. We thank T. Kodama for
contributing, 9 years ago, an unpublished private communication about
the method and essential results regarding $\pi^0$ equilibration
using the detailed balance method, and for close reading of the
manuscript and valuable comments. This work was supported  by U.S. Department of Energy Grant No.  DE-FG02-04ER41318.


\end{document}